\documentclass[12pt]{iopart}
\usepackage{graphicx}
\usepackage{amsthm}
\newtheorem{thm}{Theorem}
\begin{document}

\title{Optimal driving waveform for overdamped, adiabatic rocking ratchets}
\author{S J Lade}
\address{Nonlinear Physics Centre, Research School of Physical Sciences and Engineering, Canberra ACT 0200, Australia}
\ead{steven.lade@anu.edu.au}
\begin{abstract}
As a first step in the project of ratchet optimisation, the optimal driving waveform among a wide class of admissible functions for an overdamped, adiabatic rocking ratchet is shown to be dichotomous. `Optimum' is defined as that which achieves the maximum (or minimum negative) average particle  velocity. Implications for the design of ratchets, for example in nanotechnological transport, may follow. The main result is applicable to a general class of adiabatic responses.
\end{abstract}

\section{Introduction}
Though its history stretches back to Feynman \cite{Feynman_1963} and Smoluchowski \cite{Smoluchowski_PZ_1912}, the ratchet concept---a non-zero particle current arising due to symmetry breaking \cite{Denisov_PRE_2002,Denisov_PRA_2007} of a potential or force---has experienced a resurgence of interest in the last 15 years. Initially, the interest was motivated by the possibility of the ratchet concept explaining the operation of biological molecular motors \cite{PESKIN_BJ_1993,Spudich_Nature_1994,Julicher_RMP_1997}. Ratchets have now been realised in a wide range of physical systems \cite{Reimann_PR_2002}, with possible applications in nanotechnology \cite{Matthias_Nature_2003}, and are being explored in quantum \cite{Astumian_PT_2002} as well as classical regimes.

Most authors choose a potential and a force that preserve or break the relevant symmetries, and then investigate the dependence of the ratchet current on scalar parameters, such as damping coefficient or temperature. A few have considered a parametrised potential or force, such as a sawtooth potential with variable up-slope and down-slope gradients \cite{Sokolov_PRE_2001}, and then vary those parameters. Another such parametrisation is of the amplitudes and phase of a biharmonic force, as will be discussed in section \ref{sec:numerics}.

In this article I begin the ambitious project of \emph{optimising} the `shape', or functional form, of the symmetry-breaking force or potential \emph{over all admissible functions}, to achieve maximum ratchet current. Previous discussions of ratchet optimisation have generally been in an abstract thermodynamic sense, pointing out that maximum efficiency of a Brownian ratchet is achieved in the limit of tight coupling \cite{Gomez-Marin_PRE_2006}.

It is unlikely that one shape will be optimal for all regimes: overdamped, dissipative and Hamiltonian; fast and adiabatic driving. One specific regime is here considered, this being an overdamped, adiabatic rocking ratchet, with equation of motion
\begin{equation*}
\eta \dot x (t) = -V'(x) + f(t) + \xi(t).
\end{equation*}
This models an overdamped particle with position $x(t)$ and damping coefficient $\eta$ moving in a spatially-varying potential $V(x)$, subject to a time-varying force $f(t)$ and Brownian noise $\xi(t)$. In the adiabatic limit the average velocity of the particle in the ratchet is \cite{Reimann_PR_2002}
\begin{eqnarray}
\langle \dot{x} \rangle = \frac{1}{T_f} \int_0^{T_f} v(f(t)) dt \label{eq:adiabaticgeneral} \\
v(y) \equiv \frac{L k_B T \left[1-e^{-Ly/k_B T} \right]}{\eta \int_0^L dx \int_x^{x+L} dz \exp\left\{\left[ V(z)-V(x)-(z-x)y\right] /k_B T\right\}}, \label{eq:adiabaticratchet}
\end{eqnarray}
where the periods of the force and potential are $T_f$ and $L$, respectively, and the temperature of the Brownian noise is $k_B T$. `Adiabatic' in this context means that the frequency of the driving is slow compared to any other characteristic frequencies of the system, such as the motion of the particle in its potential $V(x)$. 

\begin{figure}
\includegraphics[width=9cm]{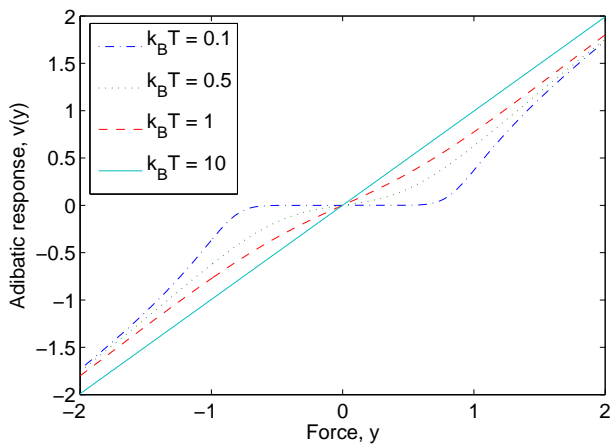}
\caption{Adiabatic response function $v(y)$ for a range of temperatures $k_B T$, with $V(x)=\cos x$ and $\eta=1$.}
\label{fig:response}
\end{figure}

The potential is here set to the symmetrical $V(x) = V\cos (2\pi x / L)$, so that an average particle velocity arises from temporal symmetry breaking only, and the remaining parameters are scaled so that $V = 1$ and $L = 2\pi$. In all calculations in this document $\eta = 1$; any other value, as can be seen from \eref{eq:adiabaticratchet}, will just scale the velocity $\langle \dot x \rangle$. Plots of the resulting function $v(y)$ are shown in Figure \ref{fig:response}. Notice that at large $y$, $v(y) \to y$; differences are only at small $y$, and that even these variations disappear as $k_B T$ becomes large.

The optimal force, as will be proved in section 2, is dichotomous, that is, it takes only two values, switching instantaneously between them. This result holds for the optimisation of any function $v(y)$ for the measure \eref{eq:adiabaticgeneral} over the admissibility criteria presented in that section. Using some more specific properties of the response function \eref{eq:adiabaticratchet}, section 3 finds further properties of the optimal force for that response function. Section 4 presents numerical illustrations and general discussion, followed by conclusions and directions for future work in section 5.

To my knowledge, only one author has previously considered the dependence of the ratchet current on the general shape of the potential or force. In a recent article, Chac\'on \cite{Chacon_JPA_2007} admirably proposed a `degree of symmetry breaking' (DSB) for each of the relevant symmetries, in an effort to quantify the connection between symmetry breaking and ratchet current. The only evidence he showed to support this measure, however, was an interpretation that for a biharmonic driving force it predicted the same optimal combination of amplitudes as the perturbation analysis discussed in section \ref{sec:numerics}. Furthermore, the measures diverge for a wide range of parameters, while any DSB, if it is to be correlated with ratchet current, should presumably remain finite. (The Cauchy principal value of the DSBs do exist, but, it can be shown, at least one is constant for all biharmonic forces, which is not much use.) DSBs will not be considered further here, except in the sense that the adiabatic response (\ref{eq:adiabaticgeneral}-\ref{eq:adiabaticratchet}) provides an indirect measure of symmetry breaking, for the current is zero when all symmetries are unbroken.

\section{The optimal force is dichotomous}
The class of admissible functions $f(t)$ over which the optimisation will take place is the set of (i) piecewise continuous functions with a finite number of discontinuities (`almost continuous'), which (ii) have period $T$, that is, $f(t + T) = f(t)$, are (iii) bounded, that is, there exists an $L$ for which $|f(t)| \le L$ for all $t$, and (iv) have zero mean, $\int_0^T f(t) dt = 0$.

\begin{thm}
If force(s) exist that optimise the current
\begin{equation*}
\langle \dot{x} \rangle(f) \equiv \int_0^T v(f(t)) dt,
\end{equation*}
then one such optimal force is dichotomous, that is, for all $t$, $f(t) = M$ or $N$, for some $M$ and $N$. The adiabatic response function $v(y)$ is assumed to have $v(0)=0$.

An `optimum' force is defined as a force $f^*$ with $\langle \dot{x} \rangle(f^*) \ge \langle \dot{x} \rangle(f)$ for maximisation or $\langle \dot{x} \rangle(f^*) \le \langle \dot{x} \rangle(f)$ for minimisation, where $f$ is any admissible force.
\end{thm}

It is trivial to see that the adiabatic response function \eref{eq:adiabaticratchet} satisfies $v(0)=0$.

\begin{proof}
First consider a constant force. This, by the constraint of zero mean, must have value zero, which gives zero velocity, and is therefore not optimal. Therefore any optimal force must take two (dichotomous) or more values.

Choose a small discretisation size $dt$ to discretise the force $f(t)$ into a series of segments $f([n + \frac{1}{2}] dt)$ on $n dt \le t < (n+1)dt$, $n = 0, \dots, T/dt-1$. Let $f_i$ refer to the discretisation of $f(t)$ at $t_i$. Consider an admissible function which is not constant and not dichotomous, that is, it takes three or more values. For a small enough $dt$ there must exist three $t_i$, $i=1,2,3$ with $f_i \ne f_j$ for $i \ne j$, that is, the force takes on at least three different values. The contribution of the three intervals to the velocity is
\begin{equation}
\label{eq:V} \left[v(f_1) + v(f_2) + v(f_3)\right] \frac{dt}{T}.
\end{equation}

\begin{figure}
\includegraphics[width=9cm]{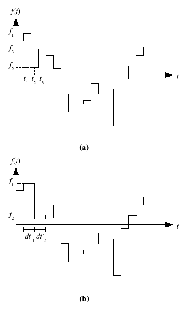}
\caption{Schematic of the method to convert a trichotomous region of the discretised force into a dichotomous region. Having chosen the three values $f_{1,2,3}$ in (a), the values of the force at those times are replaced with just $f_1$ and $f_2$ in (b). In total, the force takes values $f_{1,2}$ for total times $dt'_{1,2}$, respectively.}
\label{fig:replacement}
\end{figure}

Consider the effect on the velocity of replacing this three-valued section with just two values. Suppose $f_3$ is to be eliminated. This procedure is illustrated in Figure \ref{fig:replacement}. Let the combined lengths of the intervals at which $f(t)$ will be set to $f_1$ and $f_2$ be $dt'_1$ and $dt'_2$, respectively, where $dt'_1 + dt'_2 = 3dt$. For $\int_0^T f(t) dt$ to remain at zero,
\begin{equation}
\label{eq:zeroavenew} (f_1 + f_2 + f_3)dt = f_1 dt'_1 + f_2 dt'_2.
\end{equation}
At least one $f_i$ must be greater than the average value $(f_1 + f_2 + f_3)/3$ and at least one less than the average value. If, without loss of generality, these are labelled as $f_1$ and $f_2$, respectively, then a solution $dt'_1, dt'_2$ to \eref{eq:zeroavenew} can always be found. Using \eref{eq:zeroavenew}, the change in velocity due to the change from a trichotomous to a dichotomous region can be shown to be
\begin{equation}
\label{eq:deltav2} \Delta \langle \dot  x \rangle = (dt/T) (f_3 - f_2) (m_{21} - m_{23}),
\end{equation}
where
\begin{equation*}
m_{ij} \equiv \frac{v(f_j)-v(f_i)}{f_j-f_i}
\end{equation*}
is the gradient of the line joining $(f_i,v(f_i))$ and $(f_j,v(f_j))$.

Each of the four possible arrangements of the points $(f_i,v(f_i))$ illustrated in Figure \ref{fig:cases} will now be considered in turn.

\begin{figure}
\includegraphics[width=9cm]{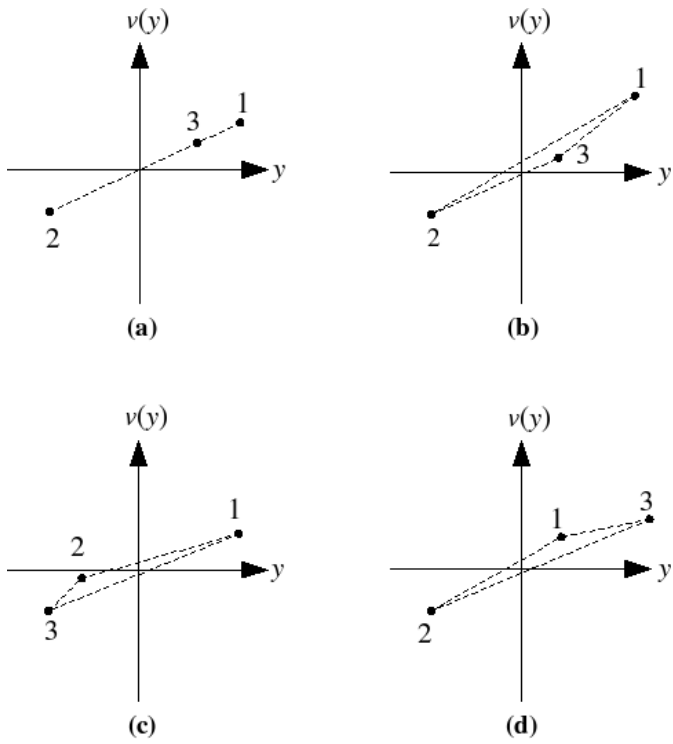}
\caption{Cases to be considered for the configuration of the points $(f_i,v(f_i)), i=1,2,3$. (a) The three points are collinear. (b) The point with intermediate $y$-value (note $y$ is the horizontal co-ordinate here) is below the line joining the other two points. (c) The point with intermediate $y$-value is above the line joining the other two points, and has $y$-value less than the average value $(f_1 + f_2 + f_3)/3$. (d) The point with intermediate $y$-value is above the line joining the other two points, and has $y$-value greater than or equal to the average value $f_\textrm{av} = (f_1 + f_2 + f_3)/3$.  The points have been numbered so that $f_1 > f_\textrm{av}$, $f_2 < f_\textrm{av}$, and removing $f_3$ in the procedure outlined in Figure \ref{fig:replacement} will increase the ratchet velocity.}
\label{fig:cases}
\end{figure}

\emph{Case illustrated by Figure \ref{fig:cases}(a).} In the collinear case there will be no change in velocity, $\Delta \langle \dot  x \rangle = 0$.

\emph{Case illustrated by Figure \ref{fig:cases}(b).} Without loss of generality the points may be labelled as shown. The middle point will therefore be `removed'. As can be seen on the figure, $m_{21} > m_{23}$, and since $f_3 > f_2$ then from \eref{eq:deltav2}, $\Delta \langle \dot  x \rangle > 0$.

\emph{Case illustrated by Figure \ref{fig:cases}(c).} Without loss of generality the points may be labelled as shown, that is, the most negative $f$ as `3'; this is the point that will be `removed'. From the figure, $m_{21} < m_{23}$, and since $f_3 < f_2$ then from \eref{eq:deltav2}, $\Delta \langle \dot  x \rangle > 0$.

\emph{Case illustrated by Figure \ref{fig:cases}(d).} Without loss of generality the most positive $f$ may be labelled as `3'. From the figure, $m_{21} > m_{23}$, and since $f_3 > f_2$ it follows that $\Delta \langle \dot  x \rangle > 0$.

Cases (c) and (d) are treated differently so that in each there exists a solution to \eref{eq:zeroavenew}, as discussed in the text below that equation.

Thus replacing the trichotomous (three-valued) region by a dichotomous one will increase the ratchet velocity or at worst keep it the same. In doing so the $N$-valued force has been replaced by an $(N-1)$-valued force. Repetition of the above procedure, at each step increasing the ratchet velocity, will end with a dichotomous force. (Some of these iterations will require operation on segments with unequal width, but the above mathematics can be easily modified to account for this.) As the width of the original discretisation $dt$ approaches zero, the above method approaches an optimisation of the original waveform.

For any admissible force, then, a dichotomous one with higher velocity can be found. Therefore an optimal force for maximum velocity, if it exists, must be dichotomous. By an analogous argument to the above, another dichotomous force must provide the most negative (minimum) velocity.
\end{proof}

The procedure in this proof will not necessarily converge to the optimum dichotomous waveform, but does always find one that gives a greater velocity than the arbitrary admissible waveform with which it began.

\section{Dichotomous optimal force for the ratchet}
This section finds further properties of the dichotomous, optimal force, assuming that 
\begin{equation}
\label{vass:gradoverv}
\mathop{\rm sign} y \; \frac{d}{dy} \frac{v(y)}{y} > 0 \textrm{   for } y \ne 0,
\end{equation}
and that $v(y)$ is continuous and smooth. The response \eref{eq:adiabaticratchet} satisfies these conditions.

The defining characteristics of a dichotomous force are the force's two values and the time spent at each value. The ratchet velocity $\langle \dot x \rangle (f)$ is clearly insensitive to time shifts. Likewise it is insensitive to `time-mixing' of the force, where it alternates many times between its values, as long as the total time spent at each value is the same, though a force with just two transitions per period is more physically realistic.

Define the dichotomous force
\begin{equation*}
f(t) = \left\{
\begin{array}{ll}
M>0, &0 \le t < Td \\
N<0, &Td \le t < T
\end{array} \right. .
\end{equation*}
Since the force must have zero mean,
\begin{equation}
\label{eq:dichotzeromean} M d + N (1 - d) = 0.
\end{equation}
The velocity can therefore be written as
\begin{equation}
\label{eq:xdotdichot}
\langle \dot x \rangle_\textrm{dichot} = v(M) d + v\left(-M \frac{d}{1-d}\right)(1-d).
\end{equation}
Setting $0 = \frac{\partial \langle \dot x \rangle}{\partial d}$, treating $M$ as constant and using constraint \eref{eq:dichotzeromean} to specify $N$ as a function of $d$ gives
\begin{equation}
\label{eq:optsq} v'(N^*) = \frac{v(M) - v(N^*)}{M - N^*},
\end{equation}
where the prime here denotes a derivative. This provides a condition for the locally extremising $N$, $N^*$. It holds for all response functions for which the optimal force is dichotomous. It can be shown that $\frac{\partial^2 \langle \dot x \rangle}{\partial d^2} \propto v''(N)$, so the type of the extremum is determined by the sign of $v''(N^*)$.

\begin{figure}
\includegraphics[width=9cm]{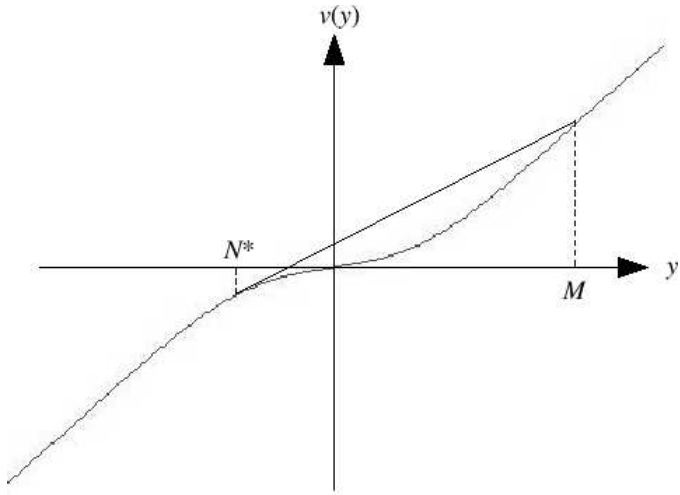}
\caption{Graphical representation of \eref{eq:optsq}.}
\label{fig:tangentsecant}
\end{figure}

The next step is to attempt optimisation with respect to $M$. Return to \eref{eq:xdotdichot}. When forming $\frac{\partial \langle \dot x \rangle}{\partial M}$ one must consider $\partial d/\partial M$; however, these terms cancel to give
\begin{equation}
\label{eq:dvdm}
\frac{\partial \langle \dot x \rangle}{\partial M} = d \left[ v'(M) - v'(N^*) \right]
\end{equation}

This quantity, as will now be shown, is greater than zero for all $M>0$. Notice first that \eref{eq:optsq} can be interpreted graphically: The tangent of $v(y)$ at $N^*$ is collinear with the secant joining $(N^*,v(N^*))$ and $(M,v(M))$, as sketched in Figure \ref{fig:tangentsecant}. An equation describing this line, to be denoted by $T$, is
\begin{equation*}
v_T(y_T) = v'(N^*) (y_T - N^*) + v(N^*).
\end{equation*}
It intercepts the vertical ($v$) axis at
\begin{equation*}
v_T(0) = -v'(N^*) N^* + v(N^*) > 0,
\end{equation*}
with the inequality following from \eref{vass:gradoverv} and $N^* < 0$. Since the line also contains $(M,v(M))$ then another equation describing it is
\begin{equation*}
v_T(y_T) = v'(N^*) (y_T - M) + v(M).
\end{equation*}
According to this equation the $v$-intercept is
\begin{equation*}
v_T(0) = -v'(N^*) M + v(M).
\end{equation*}
Suppose $v'(M) \le v'(N^*)$. Then
\begin{equation*}
v_T(0) \le -v'(M) M + v(M) < 0
\end{equation*}
with the last inequality following from \eref{vass:gradoverv} and $M>0$. But this contradicts the earlier result. Therefore it follows that in fact $v'(M) > v'(N^*)$ and by \eref{eq:dvdm} that $\frac{\partial \langle \dot x \rangle}{\partial M} > 0$ for all $M>0$. Therefore the optimum $M$, for maximum velocity, is its maximum permitted value, $L$.

In order to find the waveform that minimises the velocity the above argument may be repeated but first optimising with respect to $M$. It then follows that the optimum $N$ is the most negative possible, $-L$. If $v(y) = -v(-y)$, as does \eref{eq:adiabaticratchet}, then this force is the negative of the one just found to maximise the velocity.

By inspection of Figure \ref{fig:response}, it appears that for each $v(y)$ given by \eref{eq:adiabaticratchet} there exists exactly one $N^*$ that satisfies \eref{eq:optsq} for fixed $M$, as sketched in Figure \ref{fig:tangentsecant}. Further, it appears that $v''(N^*)$ is of the correct sign for local maximisation (in Figure \ref{fig:tangentsecant}) or minimisation. I state without proof that this is indeed the case. Since there is only one local extremum in ratchet velocity (as one varies $N$) and it is of the correct type, and $v(y)$ is smooth and continuous, it follows that the global extremum is also here.

By Theorem 1, then, there exists an optimum force for maximisation (and another for minimisation), it is dichotomous, and its characteristics are unique.

\section{Numerical work and Discussion}
\label{sec:numerics}
\begin{figure}
\includegraphics[width=9cm]{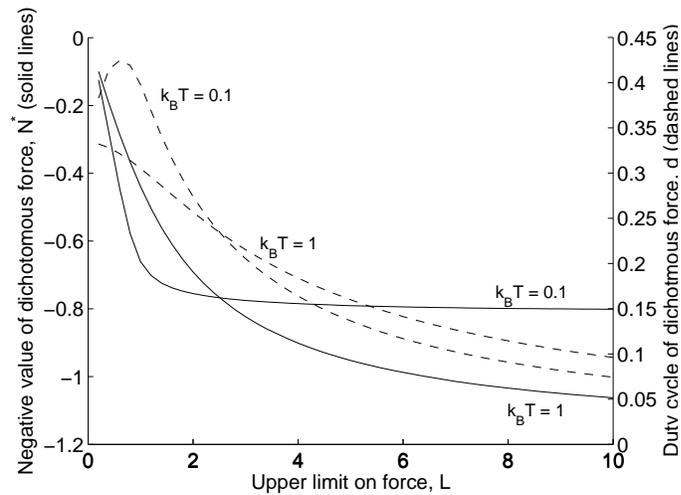}
\caption{Optimal value for the negative part of the dichotomous force (when maximising the ratchet velocity), $N^*$ (solid lines), and optimal value for the dichotomous force's duty cycle, $d$ (dashed lines), as functions of the limit $L$, for two temperatures $k_B T$.}
\label{fig:optNd}
\end{figure}

Figure \ref{fig:optNd} shows the optimal $N^*$ and $d$ at various force limits $L$ and temperatures $k_B T$. Notice that, as intuitively expected from Figure \ref{fig:tangentsecant}, for most $L$ at $k_B T = 0.1$ the optimal $N^*$ is on or close to the knee in Figure \ref{fig:response}, while at higher temperature $N^*$ changes more with $L$. As the upper limit increases, the duty cycle shows that the dichotomous, optimal force spends a smaller fraction of time at this upper value.

A common ratchet driving waveform is the biharmonic drive,
\begin{equation*}
f_\textrm{bihar}(t) = A\cos\omega t + B \cos (2\omega t + \phi).
\end{equation*}
According to perturbation analysis this will generate, in an overdamped spatially-symmetric ratchet, at the limit of small force amplitude, an average ratchet velocity that scales with $A^2 B \cos\phi$ \cite{Chacon_JPA_2007,Marchesoni_PLA_1986,Denisov_PRE_2002}. The optimal biharmonic phase is therefore $\phi = 0$ (or $\phi = \pi$, for transport in the negative direction) and, if the maximum value of the force is limited to $L$, the optimal amplitudes are $B = L/3$ and $A = 2L/3$.

\begin{figure}
\includegraphics[width=9cm]{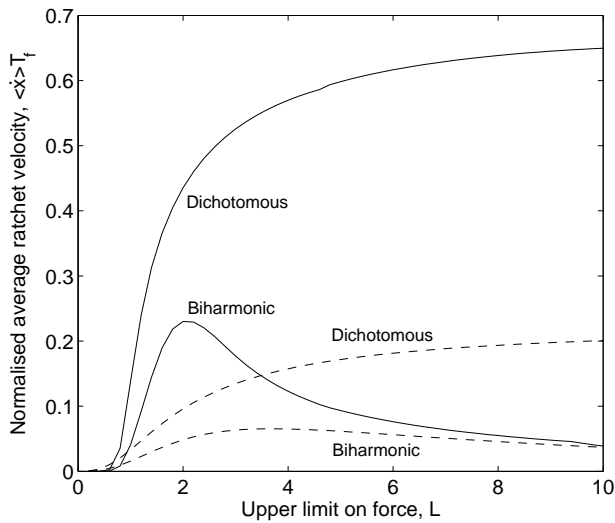}
\caption{Velocities, normalised by the force's period $T_f$, predicted by (\ref{eq:adiabaticgeneral}-\ref{eq:adiabaticratchet}) for the dichotomous optimal force and the corresponding perturbatively optimal biharmonic force with same maximum value $L$, for temperatures $k_B T = 0.1$ (solid lines) and $1$ (dashed lines).}
\label{fig:limit_velocities}
\end{figure}

The velocities predicted by (\ref{eq:adiabaticgeneral}-\ref{eq:adiabaticratchet}) are shown in Figure \ref{fig:limit_velocities} for both the optimal dichotomous force obtained previously and this perturbatively optimal biharmonic force. At small $L$ the biharmonic responses are almost as large as the dichotomous responses, as is appropriate for a perturbation analysis. For larger forces, however, the biharmonic responses are significantly less than the corresponding dichotomous responses, and even decay for very large forces. While the rates of increase of the dichotomous responses do slow, they will show no such turning point, as shown in the paragraph following \eref{eq:dvdm}.

For any experimentalist or engineer designing an overdamped ratchet close to the adiabatic limit---or even as a first guess away from the adiabatic limit---and who can freely choose the time dependence of the force up to a limiting value $\pm L$, a dichotomous force, it has been shown, will give the largest possible average ratchet velocity. The optimal characteristics of the force can be solved with the equations or approximated with the graphs given here.

That the optimal force is dichotomous is to be intuitively expected when one notes that the response \eref{eq:adiabaticgeneral}, as should be expected for an adiabatic response, depends only on the force $f$, not any of its derivatives or anti-derivatives. There is no `penalty', then, for the discontinuities in the dichotomous force. 

It is acknowledged that practical restrictions of specific experimental arrangements may in some cases more strongly influence the form of the force, for example biharmonic forces from the mixing of laser beams \cite{Schiavoni_PRL_2003}, or near-dichotomous forces from the application of pressure across a membrane \cite{Matthias_Nature_2003}.


\section{Conclusions and future work}
An optimum force for an overdamped, adiabatic ratchet, amongst the admissible forces---briefly, forces with zero mean and maximum absolute value not greater than $L$---if it exists, is dichotomous. The optimum characteristics of the dichotomous force, for maximum velocity, are (any time-shifted or time-mixed version of this force will give the same velocity)
\begin{equation*}
f^*(t) = \left\{
\begin{array}{ll}
L, &0 \le t < d \\
N^*, &d \le t < T
\end{array} \right.
\end{equation*}
where $N^* < 0$ is a solution of
\begin{equation*}
v'(N^*) = \frac{v(L) - v(N^*)}{L- N^*}
\end{equation*}
and $d$ is then given by $d = N^*/\left(N^* - L\right)$. From plots of the response \eref{eq:adiabaticratchet}, it appears that there a (unique) dichotomous optimal force does exist, though this was not formally proven. 

The main result of this article, Theorem 1, may apply to a wide range of problems. It was shown that for \emph{any} response of the form $\int_0^T v(f(t)) dt$, where $f(t)$ has a finite number of discontinuities, has zero mean and has size constrained by its maximum absolute value, the response is extremised by a dichotomous force. 

The result may be useful for engineers designing overdamped Brownian motors in the adiabatic limit, or even as a first guess away from the adiabatic limit, wishing to achieve the maximum possible average velocity from the ratchet. For large forces a dichotomous force can generate a much larger velocity than that generated by the biharmonic force optimised by perturbation analysis.

There remain many ratchet configurations in which this optimisation task could be addressed. For the case of an overdamped, adiabatic ratchet with time-symmetric force, early results indicate that the optimum potential may be close to a sawtooth waveform, which would be a pleasingly complementary result to the above. One could consider an overdamped ratchet with fast driving, from which some conclusions about the optimal waveform for all driving frequencies might be drawn. There are also the underdamped, both adiabatic and fast-driven, cases. In these analyses, more sophisticated methods such as the calculus of variations or Pontryagin's Maximum Principle \cite{Pontryagin_1962} may be useful.

\section*{References}
\bibliographystyle{unsrt}
\bibliography{ratchets}
\end{document}